\documentstyle[11pt]{article}
\begin{document}
\title{Supersymmetric $U(1)$ Gauge Field Theory \\ With Massive Gauge Field}
\author{{Ning Wu}
\thanks{ email:wuning@tofj1.ihep.ac.cn}
\\
{\small CCAST (World Lab), P.O.Box 8730, Beijing 100080, P.R.China}\\
{\small and}\\ 
{\small Division 1, Institute of High Energy Physics, P.O.Box 918-1, 
Beijing 100039, P.R.China}}
\maketitle
\vskip 0.8in

\noindent
PACS Numbers: 11.30.Pb,  12.60.J,  11.15-q,  \\
Keywords: supersymmetry, gauge symmetry, gauge field, mass \\
\vskip 0.8in

\noindent
[Abstract] A supersymmetric model with $U(1)$ gauge symmetry will be discussed in this paper. 
The model has strict $U(1)$ gauge symmetry and supersymmetry simultaneously. Besides, there is a 
massive $U(1)$ gauge field contained in the model.
\\

~~~~ According to the conventional notion, if a system has strict gauge symmetry, the mass of gauge 
field in the lagrangian must be zero \lbrack 1 \rbrack. With the development of the gauge theory, this 
notion has already been changed. In the gauge theory developed by Wu \lbrack 2 \rbrack, 
under the precondition that the system has strict gauge symmetry, the theory contains massive gauge 
bosons. In fact, this result holds not only in the usual quantum field theory, but also in the 
supersymmetric quantum field theory. In other words, we could construct a supersymmetric gauge 
field model with massive gauge bosons. Supersymmetric gauge field models have been extensively 
discussed in the literature \lbrack 3-5 \rbrack. Similar to the conventional gauge model, in all these 
supersymmetric gauge models, the masses of gauge fields is zero. In this paper, we will construct a 
gauge field model which has strict $U(1)$ gauge symmetry, supersymmetry and one massive gauge 
field.\\

~~~~ In supersymmetry, gauge field comes from vector superfield. Vector superfield is denoted 
as ${\rm V}(x, \theta, \overline{\theta})$, where $x$ is space-time coordinate, $\theta$ and 
$\overline{\theta}$ are coordinates of superspace. Vector superfield satisfies reality condition:
$$
{\rm V}(x, \theta, \overline{\theta}) = {\rm V}^{\dag}(x, \theta, \overline{\theta})
\eqno{(1)} 
$$
The most general vector superfield contains six component fields. In supersymmetric gauge field 
theory, we always select Wess-Zumino gauge. In this case, a vector superfield only contains three 
component fields. It can be expanded as:
$$
{\rm V}(x, \theta, \overline{\theta}) 
= \theta \sigma^{\mu} \overline{\theta} v_{\mu}(x)
+(\theta \theta) \overline{\theta} \, \overline{\lambda}(x)
+(\overline{\theta} \, \overline{\theta}) \theta \lambda (x)
+(\theta \theta) ( \overline{\theta} \, \overline{\theta}) D(x)
\eqno{(2)} 
$$
where $v_{\mu}(x)$ is vector field, $\lambda(x)$ is spinor field which is usually called gaugino 
field. 
In this paper, all vector superfields are required to satisfy Wess-Zumino gauge.\\

~~~~ In supersymmetry, matter fields are usually described by chiral superfields. There are two 
kinds of chiral superfields, one is left-handed chiral superfield and another is right-handed chiral 
superfield. They satisfy the following conditions
$$
\overline{D}_{\stackrel{.}{\alpha}} \Phi (x, \theta, \overline{\theta}) = 0 
\eqno{(3)} 
$$
$$
D_{\alpha} \Phi ^{\dag}(x, \theta, \overline{\theta}) = 0 , 
\eqno{(4)} 
$$
where $\Phi ^{\dag}(x, \theta, \overline{\theta}) $ is hermitian conjugation of $\Phi (x, \theta, 
\overline{\theta}) $, $D_{\alpha}$ and $\overline{D}_{\stackrel{.}{\alpha}} $ are covariant 
derivatives. \\

~~~~ In this model, we will introduce two vector superfields ${\rm V}_1 (x, \theta, 
\overline{\theta})$ and ${\rm V}_2 (x, \theta, \overline{\theta})$ , and two chiral superfields 
$\Phi_1 (x, \theta, \overline{\theta}) $ and $\Phi_2 (x, \theta, \overline{\theta}) $. Two vector 
superfields are in the Wess-Zumino gauge. They can be expanded as:
$$
{\rm V}_i (x, \theta, \overline{\theta}) 
= \theta \sigma^{\mu} \overline{\theta} v_{i \mu}(x)
+(\theta \theta) \overline{\theta} \, \overline{\lambda}_i (x)
+(\overline{\theta} \, \overline{\theta}) \theta \lambda_i  (x)
+(\theta \theta) ( \overline{\theta} \, \overline{\theta}) D_i (x) , ~~(i=1,2).
\eqno{(5)} 
$$
The chiral superfields have the following component expressions:
$$
\Phi_i (x, \theta, \overline{\theta}) 
= A_i (y) + \sqrt{2} \theta \psi _i (y) + (\theta \theta) F_i (y) , ~~~~(i=1,2)
\eqno{(6)} 
$$
where 
$$
y^{\mu}= x^{\mu} + i \theta \sigma^{\mu} \overline{\theta} .
\eqno{(7)} 
$$
In order to obtain the kinetic energy terms, let's define
$$
W_{i \alpha} = - \frac{1}{4} (\overline{D} \, \overline{D}) D_{\alpha} 
{\rm V}_i (x, \theta, \overline{\theta}) 
\eqno{(8)} 
$$
$$
\overline{W}_{i \stackrel{.}{\alpha} }
= - \frac{1}{4} (DD) \overline{D}_{\stackrel{.}{\alpha} }
{\rm V}_i (x, \theta, \overline{\theta}) 
\eqno{(9)} 
$$

~~~~ The lagrangian of the model is
$$
\begin{array}{rcl}
S & =& \int d^4 x d^4 \theta \: \lbrace
 \frac{1}{4} ( W_1^{\alpha} W_{1 \alpha}+ W_2^{\alpha} W_{2 \alpha}) 
\delta^2 (\overline{\theta})
+ \frac{1}{4} ( \overline{W}_{1 \stackrel{.}{\alpha} } \overline{W}_1^{\stackrel{.}{\alpha} }
+ \overline{W}_{2 \stackrel{.}{\alpha} } \overline{W}_2^{\stackrel{.}{\alpha} } )
\delta^2 (\theta) \\
&&+\Phi_1 ^{\dag} e^{2 e V_1} \Phi_1 + \Phi_2 ^{\dag} e^{-2 e V_1} \Phi_2  
 -m \Phi_1 \Phi_2 \delta^2 (\overline{\theta})
-m \Phi_1^{\dag} \Phi_2^{\dag} \delta^2 (\theta) \\
&& - \mu^2 ({\rm cos}\alpha V_1 +{\rm sin} \alpha V_2 )^2
\rbrace .
\end{array}
\eqno{(10)} 
$$
The supersymmetric $U(1)$ gauge transformations are:
$$
\Phi_1 (x, \theta, \overline{\theta}) 
\longrightarrow e^{- i e \Lambda} \Phi_1 (x, \theta, \overline{\theta})
\eqno{(11)} 
$$
$$
\Phi_2 (x, \theta, \overline{\theta}) 
\longrightarrow e^{ i e \Lambda} \Phi_2 (x, \theta, \overline{\theta})
\eqno{(12)} 
$$
$$
V_1 (x, \theta, \overline{\theta}) \longrightarrow
V_1 (x, \theta, \overline{\theta}) + \frac{i}{2} (\Lambda - \Lambda ^{\dag})
\eqno{(13)} 
$$
$$
V_2 (x, \theta, \overline{\theta}) \longrightarrow
V_2 (x, \theta, \overline{\theta}) - \frac{i}{2 {\rm tg} \alpha} (\Lambda - \Lambda ^{\dag}) ,
\eqno{(14)} 
$$
where $\Lambda$ satisfies the following constrains:
$$
\overline{D}_{\stackrel{.}{\alpha}} \Lambda (x, \theta, \overline{\theta}) = 0 
\eqno{(15)} 
$$
$$
D_{\alpha} \Lambda ^{\dag}(x, \theta, \overline{\theta}) = 0 .
\eqno{(16)} 
$$
Corresponding to the Wess-Zumino gauge of vector superfields, some component fields of 
$\Lambda$ should be zero. In other words, $\Lambda$'s component expression is
$$
\Lambda (x, \theta, \overline{\theta}) 
= i B(x)- \theta \sigma^{\mu} \overline{\theta} \partial _{\mu} B(x)
+ i (\theta \theta) ( \overline{\theta} \, \overline{\theta}) \partial ^{\mu} \partial_{\mu} B(x) ,
\eqno{(17)} 
$$
where $B(x)$ is a real scalar field. It is easy to prove that the action defined by eq(10) is invariant 
under the above supersymmetric $U(1)$ gauge transformations. Meanwhile, we will carry out 
integration over superspace coordinates, which equivalent to selecting the highest order component 
of 
the lagrangian. It is known that the highest order component of a superfield is invariant under 
supersymmetric transformation. So, the action defined by eq(10) has supersymmetry. \\

~~~~ Substituting component expressions into the action (10) and carrying out the integration over 
superspace coordinates, we will change the action into the following form:
$$
\begin{array}{rcl}
S&  = & \int d^4 x \: \lbrace 
2D_1^2 (x) + 2 D_2^2 (x) -  \frac{1}{4} v_{1 \mu \nu} (x) v_1^{\mu \nu} (x) 
-  \frac{1}{4} v_{2 \mu \nu} (x) v_2^{\mu \nu} (x)  \\
&& - i \lambda _1 (x) \sigma ^{\mu} \partial_{\mu} \overline{\lambda} _1 (x)
- i \lambda _2 (x) \sigma ^{\mu} \partial_{\mu} \overline{\lambda} _2 (x)  \\
&& + | F_1 (x) |^2 +i ( D_{\mu}^{\star} \overline{\psi}_1 (x) ) \overline{\sigma}^{\mu} \psi_1 (x)
+|D_{\mu} A_1 (x) |^2 + 2 e D_1 (x) |A_1 (x)|^2  \\
&& -\sqrt{2} e (\overline{\lambda}_1 (x) \overline{\psi}_1 (x) A_1 (x)
+\lambda_1(x) \psi_1(x) A_1^{\star}(x) )  \\
&& + | F_2 (x) |^2 +i ( D_{\mu} \overline{\psi}_2 (x) ) \overline{\sigma}^{\mu} \psi_2 (x)
+|D_{\mu}^{\star} A_2 (x) |^2 + 2 e D_1 (x) |A_2 (x)|^2  \\
&&+\sqrt{2} e (\overline{\lambda}_1 (x) \overline{\psi}_2 (x) A_2 (x)
+\lambda_1(x) \psi_2(x) A_2^{\star}(x) )  \\
&& +m( \psi_1 (x) \psi_2 (x) -A_1 (x) F_2 (x) -A_2 (x) F_1 (x)) \\
&&+m( \overline{\psi}_1 (x) \overline{\psi}_2 (x) 
-A_1^{\star} (x) F_2^{\star} (x) -A_2^{\star} (x) F_1^{\star} (x))   \\
&& - \frac{\mu ^2}{2} ( {\rm cos} \alpha v_{1 \mu} + {\rm sin} \alpha v_{2 \mu})
( {\rm cos} \alpha v_1^{ \mu} + {\rm sin} \alpha v_2^{ \mu})
\rbrace ,
\end{array}
\eqno{(18)} 
$$
where
$$
v_{i \mu \nu} = \partial _{\mu} v_{i \nu} - \partial_{\nu} v_{i \mu}
~~~~~~(i=1,2)
\eqno{(19)} 
$$
$$
D_{\mu}= \partial _{\mu} - i e v_{1 \mu}  ~~~~,~~~~
D_{\mu}^{\star}= \partial _{\mu} + i e v_{1 \mu}
\eqno{(20)} 
$$

~~~~ Now, let's consider the on-shell form of the aciton. In eq.(18), $F_1 (x)~,~ F_2 (x)~,~D_1 
(x)$ 
and $D_2 (x)$ are auxiliary fields which could be eliminated by applying their equations of motion. 
The equations of motion of these four auxiliary fields are:
$$
F_1^{\star} (x) = m A_2 (x)
\eqno{(21)} 
$$
$$
F_2^{\star} (x) = m A_1 (x)
\eqno{(22)} 
$$
$$
D_1 (x) = -\frac{e}{2}(| A_1 (x) |^2 - | A_2 (x) | ^2 )
\eqno{(23)} 
$$
$$
D_2 (x) = 0 .
\eqno{(24)} 
$$
Applying these four equations of motion, we could rewritten the action as:
$$
\begin{array}{rcl}
S & = & \int d^4 x \: \lbrace 
-  \frac{1}{4} v_{1 \mu \nu} (x) v_1^{\mu \nu} (x) 
-  \frac{1}{4} v_{2 \mu \nu} (x) v_2^{\mu \nu} (x)  
- i \lambda _1 (x) \sigma ^{\mu} \partial_{\mu} \overline{\lambda} _1 (x)
- i \lambda _2 (x) \sigma ^{\mu} \partial_{\mu} \overline{\lambda} _2 (x)  \\
&& +i ( D_{\mu}^{\star} \overline{\psi}_1 (x) ) \overline{\sigma}^{\mu} \psi_1 (x)
+i ( D_{\mu} \overline{\psi}_2 (x) ) \overline{\sigma}^{\mu} \psi_2 (x)
+|D_{\mu} A_1 (x) |^2 + |D_{\mu}^{\star} A_2 (x) |^2  \\
&& -\sqrt{2} e ( \overline{\lambda}_1 (x) ( \overline{\psi}_1 (x) A_1 (x)
- \overline{\psi}_2 (x) A_2 (x) )
+\lambda_1(x) ( \psi_1(x) A_1^{\star}(x) - \psi_2(x) A_2^{\star}(x) ) ) \\ 
&&+m ( \psi_1 (x) \psi_2 (x)  + \overline{\psi}_1 (x) \overline{\psi}_2 (x)  )  
 - m^2 ( | A_1 (x) |^2 + |A_2 (x)|^2 )
- \frac{e^2}{2} ( | A_1 (x) |^2 - |A_2 (x) |^2 ) ^2  \\
&& - \frac{\mu ^2}{2} ( {\rm cos} \alpha v_{1 \mu} + {\rm sin} \alpha v_{2 \mu})
( {\rm cos} \alpha v_1^{ \mu} + {\rm sin} \alpha v_2^{ \mu})
\rbrace .
\end{array}
\eqno{(25)} 
$$

~~~~ Now, we change the two-component notations into four-component notations. Define
$$
\psi (x) =\left ( 
\begin{array}{c}
\psi_ {1 \alpha} (x)\\
\overline{\psi}_2^{\stackrel{\cdot}{\alpha}} (x)
\end{array}
\right ) 
~~~~,~~~~
\overline{\psi} (x) = (
\psi_2 ^{\alpha} (x) ~,~
\overline{\psi}_{1 \stackrel{\cdot}{\alpha}} (x) )
\eqno{(26)} 
$$
$$
\lambda_1 (x) = \left ( 
\begin{array}{c}
\lambda_ {1 \alpha} (x)\\
\overline{\lambda}_1^{\stackrel{\cdot}{\alpha}} (x)
\end{array}
\right ) 
~~~~,~~~~
\overline{\lambda}_1 (x) = (
\lambda_1 ^{\alpha} (x) ~,~
\overline{\lambda}_{1 \stackrel{\cdot}{\alpha}} (x) )
\eqno{(27)} 
$$
$$
\lambda_2 (x) = \left ( 
\begin{array}{c}
\lambda_ {2 \alpha} (x)\\
\overline{\lambda}_2^{\stackrel{\cdot}{\alpha}} (x) 
\end{array}
\right ) 
~~~~,~~~~
\overline{\lambda}_2 (x) = (
\lambda_2 ^{\alpha} (x) ~,~
\overline{\lambda}_{2 \stackrel{\cdot}{\alpha}} (x) ) .
\eqno{(28)} 
$$
Then the action (25) will be changed into:
$$
\begin{array}{rcl}
S & = & \int d^4 x \: \lbrace 
-  \frac{1}{4} v_{1 \mu \nu} (x) v_1^{\mu \nu} (x) 
-  \frac{1}{4} v_{2 \mu \nu} (x) v_2^{\mu \nu} (x)  
- \frac{i}{2} \overline{\lambda} _1 (x) \gamma ^{\mu} \partial_{\mu} \lambda _1 (x)
- \frac{i}{2} \overline{\lambda} _2 (x) \gamma ^{\mu} \partial_{\mu} \lambda _2 (x)  \\
&& - \overline{\psi} ( i \gamma^{\mu} D_{\mu} - m ) \psi 
+|D_{\mu} A_1 (x) |^2 + |D_{\mu}^{\star} A_2 (x) |^2  
 - m^2 ( | A_1 (x) |^2 + |A_2 (x)|^2 )   \\
&& - \frac{e}{\sqrt{2}}  ( \overline{\psi} (1 + \gamma ^5) \lambda_1(x) A_1 (x)
 - \overline{\psi} (x)  (1 - \gamma ^5) \lambda_1(x) A_2^{\star} (x)  \\
&&+ \overline{\lambda}_1 (x) (1 - \gamma ^5) \psi (x) A_1^{\star} (x)
- \overline{\lambda}_1 (x) (1 + \gamma ^5) \psi (x) A_2 (x))   \\
&& - \frac{e^2}{2} ( | A_1 (x) |^2 - |A_2 (x) |^2 ) ^2  
 - \frac{\mu ^2}{2} ( {\rm cos} \alpha v_{1 \mu} + {\rm sin} \alpha v_{2 \mu})
( {\rm cos} \alpha v_1^{ \mu} + {\rm sin} \alpha v_2^{ \mu})
\rbrace .
\end{array}
\eqno{(29)} 
$$

~~~~ Please note that $v_{1 \mu}$ and $v_{2 \mu}$ are not eigenvectors of mass matrix, so we 
make the following transformations:
$$
F_{\mu}={\rm cos}\alpha v_{1 \mu} + {\rm sin}\alpha v_{2 \mu}
\eqno{(30)} 
$$
$$
F_{2 \mu}= - {\rm sin}\alpha v_{1 \mu}+{\rm cos}\alpha v_{2 \mu} .
\eqno{(31)} 
$$
Correspondingly, the gauge covariant derivative becomes
$$
D_{ \mu}= \partial _{\mu} - i e {\rm cos}\alpha F_{\mu} + i e {\rm sin}\alpha F_{2 \mu}.
\eqno{(32)} 
$$
After these changes, the on-shell action becomes:
$$
\begin{array}{rcl}
S & = & \int d^4 x \: \lbrace 
-  \frac{1}{4} F_{ \mu \nu} (x) F^{\mu \nu} (x) 
- \frac{\mu ^2}{2} F_{\mu} F^{\mu}
-  \frac{1}{4} F_{2 \mu \nu} (x) F_2^{\mu \nu} (x)  \\
&& - \frac{i}{2} \overline{\lambda} _1 (x) \gamma ^{\mu} \partial_{\mu} \lambda _1 (x)
- \frac{i}{2} \overline{\lambda} _2 (x) \gamma ^{\mu} \partial_{\mu} \lambda _2 (x)  
 - \overline{\psi} ( i \gamma^{\mu} D_{\mu} - m ) \psi   \\
&& +|D_{\mu} A_1 (x) |^2 + |D_{\mu}^{\star} A_2 (x) |^2  
 - m^2 ( | A_1 (x) |^2 + |A_2 (x)|^2 )
- \frac{e^2}{2} ( | A_1 (x) |^2 - |A_2 (x) |^2 ) ^2  \\
&& - \frac{e}{\sqrt{2}}  ( \overline{\psi}(x)  (1 + \gamma ^5) \lambda_1(x) A_1 (x)
 - \overline{\psi} (x)  (1 - \gamma ^5) \lambda_1(x) A_2^{\star} (x)  \\
&&+ \overline{\lambda}_1 (x) (1 - \gamma ^5) \psi (x) A_1^{\star} (x)
- \overline{\lambda}_1 (x) (1 + \gamma ^5) \psi (x) A_2 (x) )  
\rbrace ,
\end{array}
\eqno{(33)} 
$$
where ,
$$
F_{ \mu \nu} = \partial _{\mu} F_{ \nu} - \partial_{\nu} F_{ \mu}
\eqno{(34)} 
$$
$$
F_{2 \mu \nu} = \partial _{\mu} F_{2 \nu} - \partial_{\nu} F_{2 \mu} .
\eqno{(35)} 
$$

~~~~ From on-shell action (33), we know that there exist two gauge fields $F_{\mu} (x)$ and $F_{2 
\mu} (x)$ in the theory. And $F_{\mu}(x)$ is massive gauge field whose mass is $\mu$, $F_{2 
\mu}(x)$ is massless gauge field. Correspondingly, there are two massless gaugino fields 
$\lambda_1(x)$ and $\lambda_2(x)$ in the theory. The matter fields in the theory are one spinor field 
$\psi (x)$ and two complex scalar fields $A_1 (x)$ and $A_2 (x)$ whose masses are $m$. \\

~~~~ In the above discussions, only vector superfield ${\rm V}_1 (x, \theta, \overline{\theta})$ 
couples to matter fields, vector superfield ${\rm V}_2 (x, \theta, \overline{\theta})$ does not 
couple 
to matter fields directly. In fact, we could let both two vector superfields  couple to matter fields. In 
this case, the action should be defined as:
$$
\begin{array}{rcl}
S & =& \int d^4 x d^4 \theta \: \lbrace
 \frac{1}{4} ( W_1^{\alpha} W_{1 \alpha}+ W_2^{\alpha} W_{2 \alpha}) 
\delta^2 (\overline{\theta})
+ \frac{1}{4} ( \overline{W}_{1 \stackrel{.}{\alpha} } \overline{W}_1^{\stackrel{.}{\alpha} }
+ \overline{W}_{2 \stackrel{.}{\alpha} } \overline{W}_2^{\stackrel{.}{\alpha} } )
\delta^2 (\theta) \\
&&+\Phi_1 ^{\dag} e^{2 e ({\rm cos}^2 \phi V_1-{\rm tg}\alpha {\rm sin}^2 \phi V_2) } \Phi_1 
+ \Phi_2 ^{\dag} e^{-2 e ({\rm cos}^2 \phi V_1-{\rm tg}\alpha {\rm sin}^2 \phi V_2 )} \Phi_2  \\
&& -m \Phi_1 \Phi_2 \delta^2 (\overline{\theta})  
-m \Phi_1^{\dag} \Phi_2^{\dag} \delta^2 (\theta) 
 - \mu^2 ({\rm cos}\alpha V_1 +{\rm sin} \alpha V_2 )^2
\rbrace .
\end{array}
\eqno{(36)} 
$$
The above action has both supersymmetry and $U(1)$ gauge symmetry. The corresponding on-shell 
action is
$$
\begin{array}{rcl}
S & = & \int d^4 x \: \lbrace 
-  \frac{1}{4} F_{ \mu \nu} (x) F^{\mu \nu} (x) 
- \frac{\mu ^2}{2} F_{\mu} F^{\mu}
-  \frac{1}{4} F_{2 \mu \nu} (x) F_2^{\mu \nu} (x)  \\
&& - \frac{i}{2} \overline{\lambda} _1 (x) \gamma ^{\mu} \partial_{\mu} \lambda _1 (x)
- \frac{i}{2} \overline{\lambda} _2 (x) \gamma ^{\mu} \partial_{\mu} \lambda _2 (x)  
 - \overline{\psi} ( i \gamma^{\mu} D'_{\mu} - m ) \psi   \\
&& +|D'_{\mu} A_1 (x) |^2 + |D_{\mu}^{\prime \star} A_2 (x) |^2  
 - m^2 ( | A_1 (x) |^2 + |A_2 (x)|^2 )  \\
&& - \frac{e^2}{2} ( {\rm cos}^4 \phi + {\rm tg}^2 \alpha
{\rm sin}^4 \phi ) ( | A_1 (x) |^2 - |A_2 (x) |^2 ) ^2  \\
&& - \frac{e}{\sqrt{2}}  ( \overline{\psi}(x) (1 + \gamma ^5) 
( {\rm cos}^2 \phi \lambda_1(x) - {\rm tg} \alpha {\rm sin}^2 \phi \lambda_2 (x)) A_1 (x) \\
&& - \overline{\psi} (x)  (1 - \gamma ^5) 
( {\rm cos}^2 \phi \lambda_1(x) - {\rm tg} \alpha {\rm sin}^2 \phi \lambda_2 (x)) A_2^{\star} (x) \\
&& + ( {\rm cos}^2 \phi \overline{\lambda}_1(x) 
- {\rm tg} \alpha {\rm sin}^2 \phi \overline{\lambda}_2 (x))(1 - \gamma ^5) \psi (x) A_1^{\star} (x) 
\\
&&- ( {\rm cos}^2 \phi \overline{\lambda}_1(x) 
- {\rm tg} \alpha {\rm sin}^2 \phi \overline{\lambda}_2 (x))(1 + \gamma ^5) \psi (x) A_2 (x) )  
\rbrace ,
\end{array}
\eqno{(37)} 
$$
where,
$$
D'_{ \mu}= \partial _{\mu} - i e \frac{{\rm cos}^2 \phi - {\rm sin}^2 \alpha}{{\rm cos}\alpha } 
F_{\mu} 
+ i e {\rm sin}\alpha F_{2 \mu}.
\eqno{(38)} 
$$
It is clearly seen from eq(39) that gaugino field $\lambda_2 (x)$ couples to matter fields directly. 
The coupling between vector superfield ${\rm V}_2 (x, \theta, \overline{\theta})$ and matter fields 
does not affect the masses of particles, only affects the dynamical behavior of the system. \\

~~~~ From above discussions, we see that, after introducing another vector superfield, we could 
make one of the gauge field massive and keep another gauge field massless. But both gaugino fields 
are massless. So, in this model, the various components of the supermultiplet do not have the same 
masses. But, the various components of matter fields supermultiplet have the same masses.  \\

\section*{Reference:}
\begin{description}
\item[\lbrack 1 \rbrack]  C.N.Yang, R.L.Mills, Phys Rev {\bf 96} (1954) 191
\item[\lbrack 2 \rbrack]  Ning Wu, Gauge Field Model With Massive Gauge Bosons, hep-ph/9802236
\item[\lbrack 3 \rbrack]  J.Wess and B.Zumino, Nucl. Phys. {\bf  B78} (1974) 1
\item[\lbrack 4 \rbrack]  S.Ferrara and B.Zumino, Nucl.Phys. {\bf  B79} (1974)  413
\item[\lbrack 5 \rbrack]  A.Salam and J.Strathdee, Phys. Lett.  {\bf  51B} (1974) 353
\end{description}

\end{document}